# New dynamics of energy use and $CO_2$ emissions in China

Zhu Liu[1,2], Bo Zheng[3,4], Qiang Zhang[1]

1 Department of Earth System Science, Tsinghua University, Beijing 100084, China

2 Tydnall Centre for Climate Change Research, School of International Development, University of East Anglia, Norwich NR4 7TJ, UK

3 School of Environment, Tsinghua University, Beijing 100084, China

4 Laboratoire des Sciences du Climat et de l'Environnement, CEA-CNRS-UVSQ, UMR 8212, Gif-sur-Yvette, France

**[Summary]**

Global achievement of climate change mitigation will heavy reply on how much of $CO_2$ emission has and will be released by China. After rapid growth of emissions during last decades, China's $CO_2$ emissions declined since 2014[1] that driven by decreased coal consumption, suggesting a possible peak of China's coal consumption and $CO_2$ emissions[2]. Here, by combining a updated methodology and underlying data from different sources, we reported the soaring 5.5% (range: +2.5% to +8.5% for one sigma) increase of China's $CO_2$ emissions in 2018 compared to 2017, suggesting China's $CO_2$ is not yet to peak and leaving a big uncertain to whether China's emission will continue to rise in the future. Although our best estimate of total emission (9.9Gt $CO_2$ in 2018) is lower than international agencies[3-6] in the same year, the results show robust on a record-high energy consumption and total $CO_2$ emission in 2018. During 2014-2016, China's energy intensity (energy consumption per unit of GDP) and total $CO_2$ emissions has decreased driven by energy and economic structure optimization. However, the decrease in emissions is now offset by stimulates of heavy industry production under economic downturn that driving coal consumption (+5% in 2018), as well as the surging of natural gas consumption (+18% in 2018) due to the government led "coal-to-gas" energy transition to reduce local air pollutions. Timing policy and actions are urgent needed to address on these new drivers to turn down the total emission growth trend.

**Main text**

**1. China's global significance**

The trends of global $CO_2$ emissions from fossil fuel combustion and cement production process is heavily rely on the emissions from China. China now account for almost 30% of global total $CO_2$ emissions from fossil fuel consumption and cement production, and take account for 80% of the new increased emission during 2002-2010[7,8]. Whether China have peaked its total fossil energy consumption and $CO_2$ emissions is also key for achieving the tipping point of global total $CO_2$ emissions[9,10]. For instance, $CO_2$ emissions in China has been reported a decline in 2014-2016, that result in only 0.2% of the increasing of global total emissions in 2016, considerably lower than its soaring growth rate (more than 3% $yr^{-1}$) in the period of 2000-2015[8]. The 3.5% regrowth of China's $CO_2$ in year 2017, on the contrast, driven a 1.6% of the increase in global total $CO_2$ emission from fossil fuel combustion and cement production in the same year[8]. The energy statistics show the declines of total coal consumption since 2014, and several previous studies suggested a peak of China's $CO_2$ emission and coal consumption around year 2013 and 2014[1,2,11], implies a possible tipping point of the decline of global total emissions.

## 2. Rebound of China's Energy Consumption and CO2 emissions in 2018

By adopting the updated methodology (See Methods) based on the *in situ* monitoring and investigation of China's energy and CO2 emission data for decades, here we reported the significant +5.5% (range: +2.5% to +8.5% for one sigma) rebound of the total $CO_2$ emission in 2018 that compared to 2017. We compiled China's energy consumption and the $CO_2$ emissions from different estimates between 2000 and 2018, including ourselves, that list in Figure 1, China's $CO_2$ emissions slightly decreased after reaching a temporary maximum value around 2013. Since 2017, China's coal use and total $CO_2$ emissions started to rebound (Figure 1a). All the datasets that extend to cover the year of 2017 suggest that China's emissions have risen up in 2017 compared to 2016, with growing rates spanning from +1.2 to +1.7% in different estimates. In 2018, China's estimated emissions rise substantially from our results according to the first ten months of energy statistics[12]. We estimated growth rate of +5.5% (range: 2.5% to 8.5%) in 2018, which is considerably higher compared to the growth in 2017, and comparable to the soaring growth rate (5% in average) in the first decade of 2000s, in which the period China has fast emission growth (Figure 1c).

Different estimates present consistent results that China's energy consumption and $CO_2$ emissions in 2018 could have surpassed the peak point so far in 2013-2014 or in 2017 dependent on the data sources (Figure 1b). The data of UN[13] and BP[4] show higher numbers in 2017 and the CDIAC shows that the 2017 emission is slightly lower than the 2013 level by -0.2%. Although emission estimates for 2018 by UN, BP, EDGAR[6] and CDIAC have not been released by yet, these emission estimates are projected to a record high emission number in 2018, because the first ten months of energy consumption in 2018 is already around 4.5% higher than the total amount of energy consumption in whole year 2017.

The rebound of the energy consumption and associated emission in 2018 is mainly contributed by coal use +4.8% (range: 1.8% to 7.8%), oil +5.6% (range: 1.3% to 9.9%), and natural gas +17.4% (range: 14.2% to 20.6%). Coal consumption as the major fuel source in China (68% of total energy supply) has reached its fast increase since 2012. However, our results suggested that coal consumption still haven't surpass the record high point in 2013. The growth rate of natural gas consumption reached an unprecedented 17.4% in 2018. Overall the year-by-year growth rate of primary energy show that China is now in another period of fossil fuel expansion, following the previous ones in 2008 and 2012 that triggered by government's economic stimuluses.

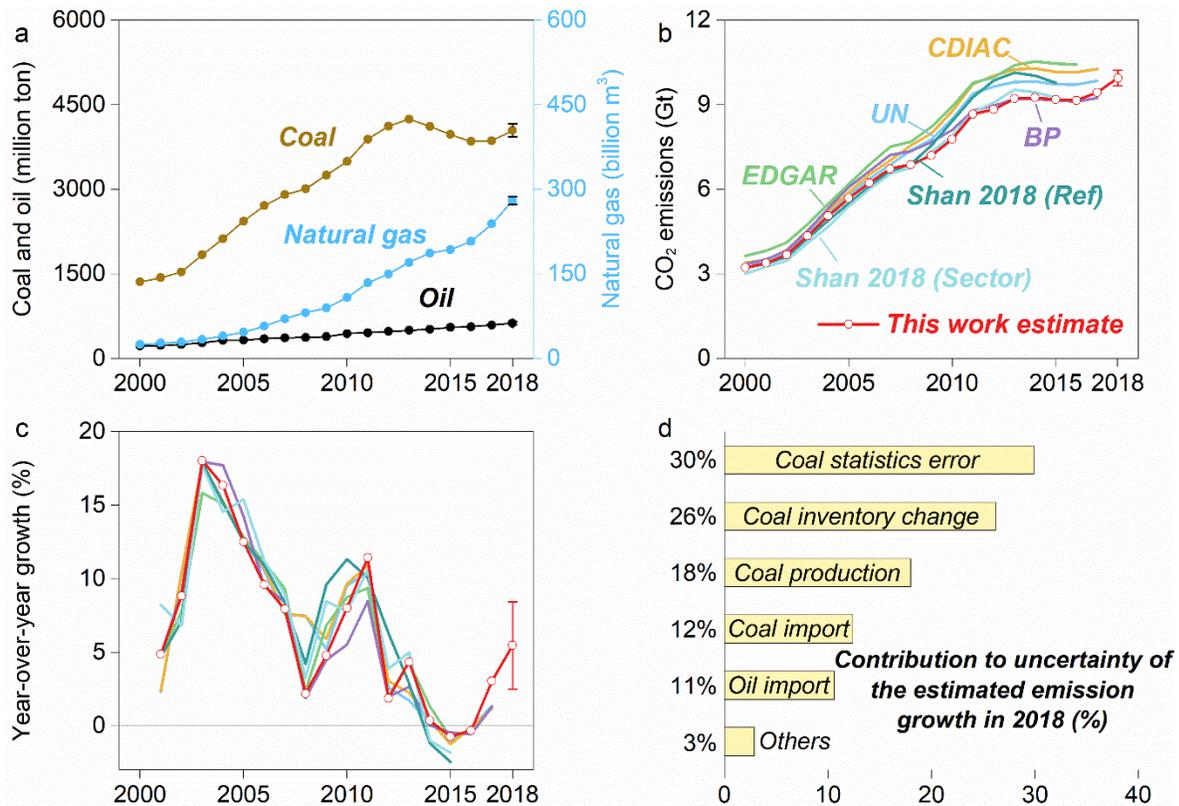

**Figure 1 Trends of China's Energy Consumption and $CO_2$ emissions (2000-2018).** Figure 1a shows the apparent energy consumption for coal, oil and gas in 2000-2018. Figure 1b shows the $CO_2$ emission estimates by different data sources in 2000-2018, Figure 1c shows the Year-over-year growth rate the $CO_2$ emission estimates. Figure 1d is the contributions of the uncertainty of total $CO_2$ increase in 2018 by different sources.

### 3. Uncertainty and projection of China's energy consumption and CO2 emissions in 2018

Data uncertainty is the key challenge for understanding China's energy and $CO_2$ emission. Although multiple data sources suggested similar trends and new record high $CO_2$ emission in 2018, the $CO_2$ emission estimated could differ by 20% by different emission sources. Data differ considerably from: energy statistics (e.g., the amount of fuel burnt or energy produced); the emission factors reflected by the heating values and the carbon content of the combusted fuels (vary by type and quality of fuels); the combustion efficiency shown as oxidization rates of the fuels as well as differences in cement emissions that associated with the production of CaCO3 with cement production. In fact, China have shown the big variations in both statistics and investigated data.

For $CO_2$ emission estimates before 2018 that the calculation is based on historical data, we adopted the uncertainty range (± 7.4%) from Monte Carlo analysis by Liu et al.[14], which considered China's data variations on statics and investigated data and covered the uncertainties from all these sources. In this analysis we further carefully address the key sources of uncertainty from:

(a) Energy Data

China has been questioned about its reliability and precise on the statistics and data reporting for a long time[15-17]. Previously 20% difference of energy consumption had reported[18,19] at the provincial level compared to the national level. In addition, Chinese government revises the energy statistics data several times after the initial publication[20]. The government reported the adjustment of amount by double-checking after years of initial publication, which suggests the improvement of data reliability after government revision. Overall the revised energy statistics result in the upscale of the total coal consumption in year 2002-2013[20].

Due to the frequent revision of China's government reported national energy consumption[21], we adopted apparent energy consumption approach (See Methods) to re-calculate China total energy consumption. The apparent energy consumption based on fuel production and trade statistics are more reliable and consistent than data of final energy consumption[14]. Comparing the more 20% difference in the data of final energy consumption[18,22], the statistics error of data on production and trade for primary energy is within 2% (Figure 2), in addition, the re-calculated China's energy consumption based on apparent energy consumption approach appears to be more close to the energy consumption data after government revision[20]. The growth rate of our recalculated energy consumption matches with the growth of industrial production which also indicating the data robustness. Overall, China's energy consumption based on apparent energy consumption approach has -3% to 2% difference (Figure 2e) with the government reported national energy consumption after government revision (in energy unit, i.e. joule).

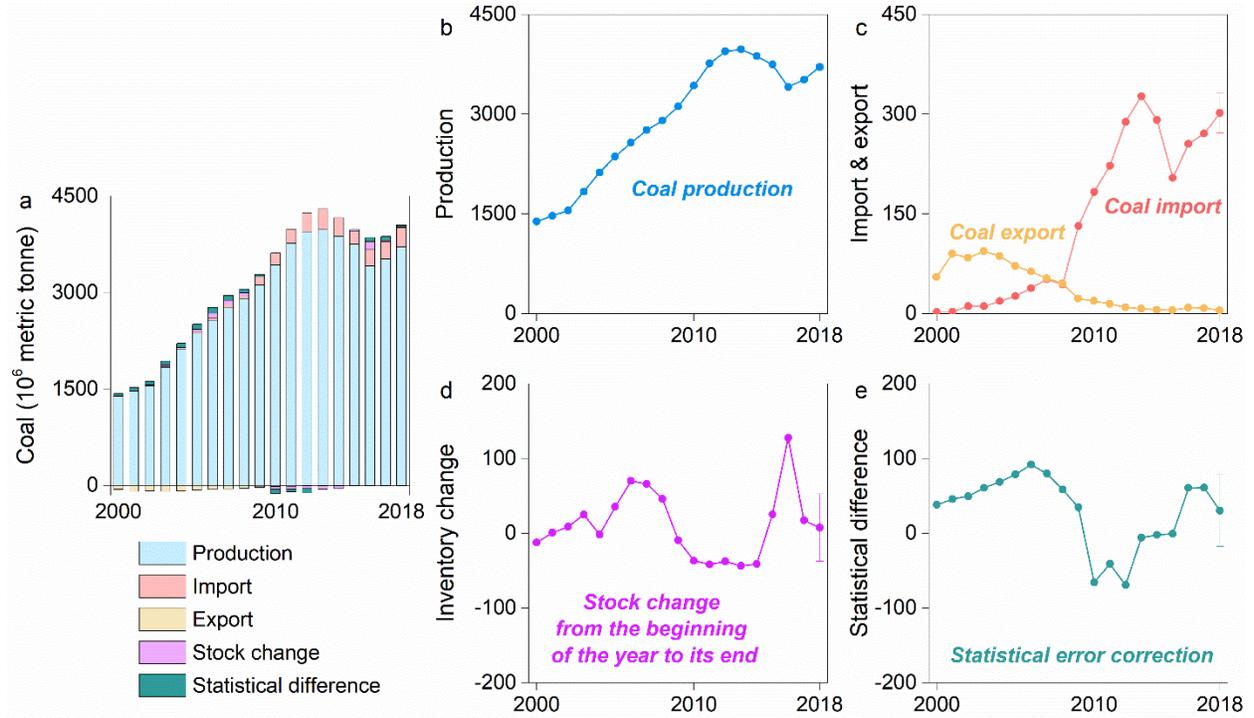

**Figure 2 Apparent energy consumption data for coal (2000-2018)**

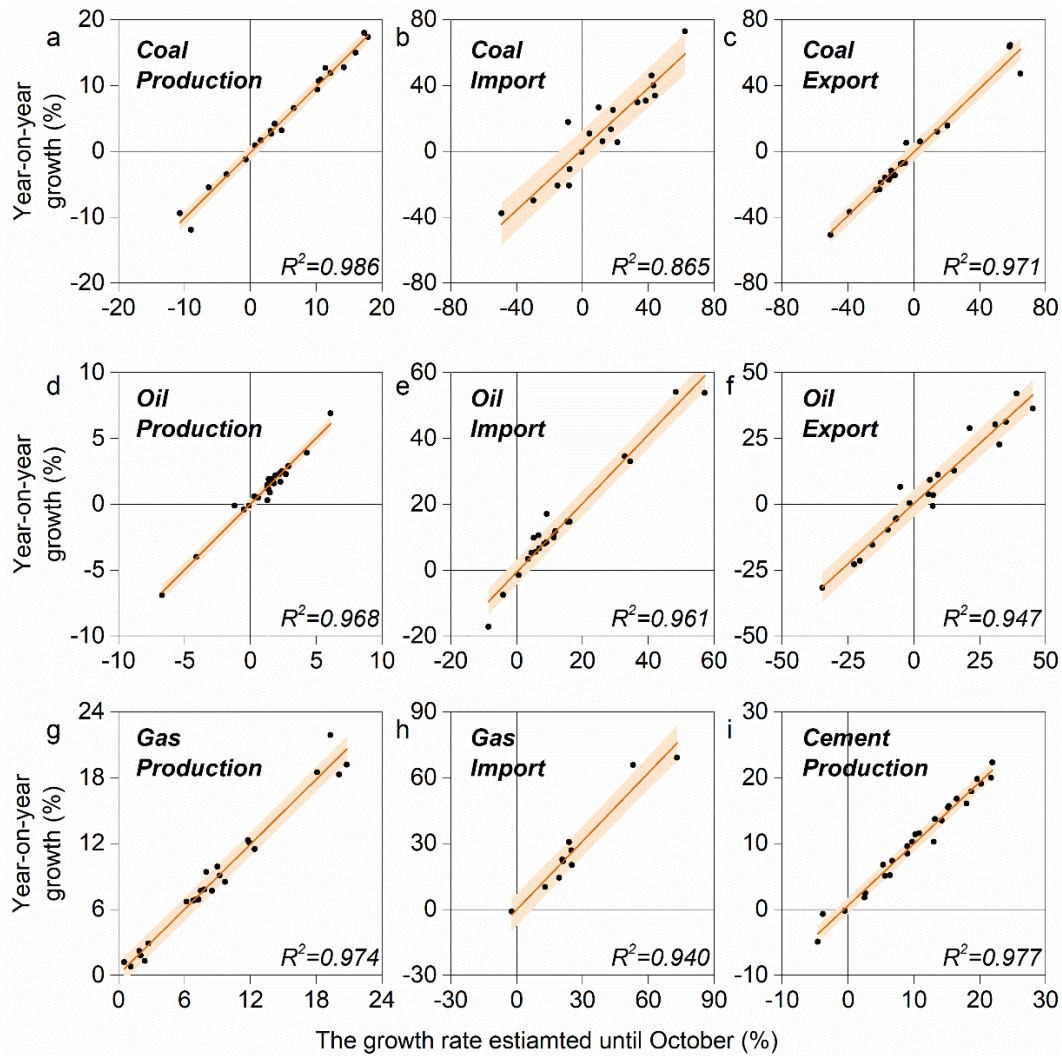

**Figure 3 Correlation between the growth rates of energy production, import, and export estimated from the first ten months with those from the whole year between 1990 and 2017. The x value is certain year's growth rate that calculated based on the first 10 months of the year, the value is the growth rate for whole year. The consistence of the results show robust about the calculation of one year's growth rate based one the rate of first 10months.**

The uncertainty of the projected data in 2018 is critical to certain whether there is a big rebound of energy and CO2 emissions. Through the apparent energy consumption approach we projected energy consumption for the whole year in 2018 based on the first 10months data[12](the projections based on 9 months data listed in SI Table 4) , in order to offset the two years lag for releasing official energy consumption data[21]. To quantify uncertainty in using 10 months data to represent the whole year's growth, we collected monthly data of energy production, import, and export for 27 years between 1990 and 2017, and used linear regression models to correlate the year-on-year growth rates calculated from the cumulative sum of the first ten months with those calculated from the whole year total. We calculated the 68% prediction interval of these linear regression models (shading in Fig. 3) and used them to reflect the uncertainty involved when using the first 10 months data to represent the whole year's variation. The squared correlation coefficients are within the range of 0.865 and 0.986, representing that using the data of the first ten months can explain 86.5% to 98.6% of the variation of the whole year (Fig. 3), while the remaining variation not covered yet reflect the uncertainty in the evolution of China's economics in November and December. The stock changes of coal, oil, and natural gas lack of enough monthly statistics data, therefore we used the standard deviation of the annual data between 2000 and 2017 as the one-sigma uncertainty. Overall the whole year projections by using 10months data show consistent and robust during 1990-2017.

(b) Emission factors

The $CO_2$ emission is calculated by the energy consumption multiplied by the associated emission factors[23], in which we used updated emission factors to deliver the best estimates. The emission factors are composed by heating values (the unit energy per unit mass fuel consumed, e.g. joule per gram coal combustion) and the carbon contents (the unit of mass carbon per unit energy, e.g. Gg CO2·TJ$^{-1}$).

As major fuel supply, China's coal has its unique low heating values (20.95GJ·T$^{-1}$) comparing with the global average heating value(29.3GJ·T$^{-1}$) provided by UN[13], as well as the heating value of other major coal production countries (SI Table 2).

In addition, the heating values varies dependent on the coal mix and thus varies over time. In this study we adopted the dynamic time-dependent heating values (Figure 4) for $CO_2$ emission estimates. The average heating value reported by National Bureau of Energy Statistics in different years is in concert with the national average heating value by sampling test in 2012[14]. Heating value of coal is the average number weighted by their consumption in specific years. In general, for years with increasing consumption and coal supply shortage, the coal with less quality and heating value will have more consumption due to profit consideration, in which result in a lower level of average heating value. For example, the coal heating values show decrease in 2005-2011 that accompanied with fast coal consumption, however, recent rebound of coal consumption is associated with increase of the national average heating values, which result in even higher CO2 emissions.

The value of carbon content we used (26.59TC·TJ$^{-1}$, 1-sigma range ± 0.3%) is based on sampling investigation, which the variation is within 2% of the IPCC default value(25.9TC·TJ$^{-1}$) and show little different in literatures.

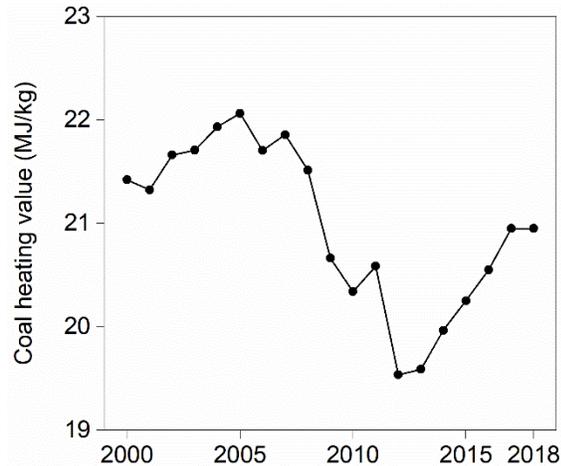

**Figure 4 The trajectory of national coal's heating values in 2000-2018**

In emission estimates, we considered the China specific oxidation rate that representing the technology for fuel combustion. Emission estimated by EGDAR, BP, World Bank and EIA do not considered the oxidation rate (assuming full combustion with 100% oxidation) for coal consumption (SI Table 3). The oxidation is based on tests for 135 different combustion technologies conducted by Chinese UNFCCC report, which report oxidation rate (94%) for the China's emission inventory in 2005 and 2012. Our oxidation has 2% less because the consideration of the mass loss from coal production to consumption. Notably with the technology development the oxidation rate is expected to increase. However so for there is no estimates based on time dependent oxidation rate, and China's UNFCCC is the only source so far reporting China specific average oxidation rate. In addition the possible increase of the oxidation rate in implies a higher $CO_2$ emission in recent years, which is in consistent with our conclusions.

In all, we have adopted China's specific emission factors for $CO_2$ emissions from fossil fuel combustion based by decades monitoring and investigations. We also adopted China specific emission factor[24,25] for cement production process, which the value is significantly lower than IPCC default value due to the less clinker proportion of Chinese cement. China's low emission factor for cement production and its explanation has been sufficiently discussed[24-26].

## 4. Driving factors and Projections

It's unclear that the rise of emission will continue as a long-term trend or is just a temporary fluctuates. China is still on track with its pledged plan to reduce the $CO_2$ emission intensity (emission per unit GDP) by 65% by 2030 comparing with the level in 2005, and the peak of total $CO_2$ emission by 2030[27,28]. China's emission intensity is playing its central role in offsetting the total emissions, plus rapid change of industrial and energy structure with more service and renewables taking the shares that significantly reduced China's total $CO_2$ during 2014-2016. However, recent rebound of energy and emission in 2018 alarm us the recent changes of the driving factors to the total emissions.

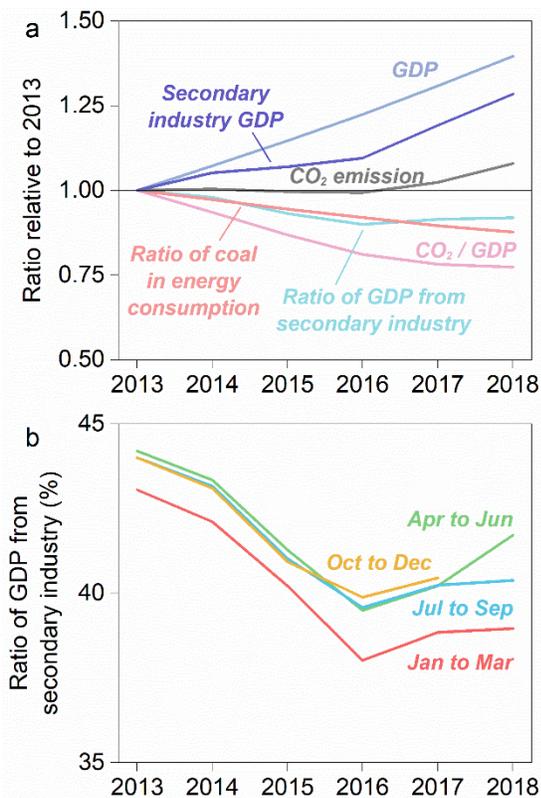

**Figure 5 trends in the driving factors of China's energy and emission**

Figure 5 show a slowdown of the $CO_2$ intensity ($CO_2$/GDP, red line) since 2017, the main driver that offset the emission increase. Although the share of coal in total fuel supply is decreasing since 2013, such trend is concert with the enlarge of total GDP volume. Moreover, the total amount of GDP from secondary industry (manufacturing) growth faster since 2016, as well as the share of secondary industry in total economy that rebound since 2016, implies the whole economy tends to be more energy intensive thereafter. Especially, the share of secondary industry in total economy (Figure 5b) show apparent rebound trend from Spring to Winter since 2016. As results, total $CO_2$ emission start to increase in 2016. We proposed several reasons as explanation:

First, China's economy slow down the growth pace to 6.5% in the first three quarters of 2018, the lowest growth level in past 10 years. The US-China trade dispute further diluted the perspective of economic growth (one fourth of China's emission is associated with exports). Central government led investment on infrastructure as stimulus for keep economic growth has driven the production of heavy industry and energy consumption. Figure 6 show the rebound of industrial products in 2018 with the highest growth rate since 2008 (Figure 6a), which the year 2008 China launched 4,000 billion RMB investment plan for infrastructure construction to counter the world economic crises. The growth rate of industrial products show around 5% increase (Figure 6b) of industrial product such as steel, iron and cement in 2018 when comparing with 2017 (for same first 10 months).

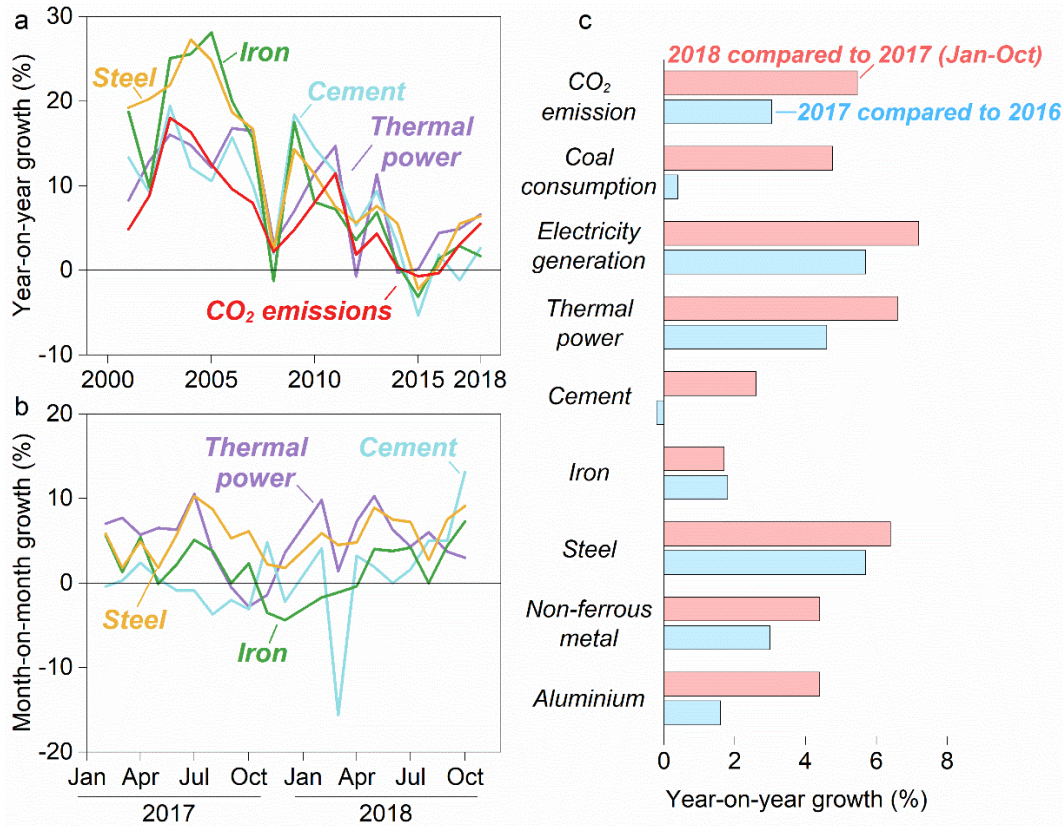

**Figure 6 Growth rates of industrial products.**

Furthermore, against central government policy to control the capacity of heavy industries, local government expanded debit and investment of the infrastructure construction to highlighting economic performance. For example, there was a surge of power plants construction due to the regulatory devolution of power plants permitting from central to provincial authorities during 2014-2016, the reported power plants construction booming is associated with more than 250Gigawatts (GW) of new capacity development in China, comparable to the entire U.S. coal fleet (266 GW) [29].

Finally, consumption of oil and natural gas is unprecedented since 2017. China is now the world largest vehicle market that driving the oil consumption. Besides, natural gas consumption has surging by national wide coal-to-gas facility renovation, since 2013 Chinese government issued serious regulations to release the local air pollution, and one key action is to replace coal by natural gas for heating in cities. The expansion of natural gas facility is so quick that result in the short natural gas supply in residential heating in winter of 2016 and 2017.

There are still big challenges for China to peak its $CO_2$ emissions, and further actions are urgent needed to turndown the emission growth trend as soon as possible.

## Methods

### 1. Carbon emission from fossil fuel combustion and the cement process

Carbon emissions are calculated by using the physical units of fossil fuel consumption multiplied by an emission factor (*EF*).

$$\text{Emission} = \text{energy consumption data} \times \text{emission factor } (EF) \quad (1)$$

If data on sectorial and fuel-specific activity data and EF are available, total emission can be calculated by:

$$\text{Emission} = \sum\sum\sum (Sectoral\ energy\ data_{i,j,k} \times EF_{i,j,k}) \quad (2)$$

Where i is an index for fuel types, j for sectors, and k for technology type. Sectoral energy consumption data is the energy consumption by individual sector that measured in physical units (t fuel).

*EF* can be further separated into net heating value of each fuel "*v*", the energy obtained per unit of fuel (TJ per t fuel), carbon content "*c*" (t C $TJ^{-1}$ fuel) and oxidization rate "*o*" the fraction (in %) of fuel oxidized during combustion and emitted to the atmosphere. The value of *v, c* and *o* are specific of fuel types, sectors and technologies.

$$\text{Emission} = \sum\sum\sum (Energy\ consumption\ data_{i,j,k} \times v_{i,j,k} \times c_{i,j,k} \times o_{i,j,k}) \quad (3)$$

Emissions from the cement manufacturing process are estimated as:

$$Emission_{cement} = \text{Production data}_{cement} \times EF_{cement} \quad (4)$$

$EF_{cement}$ is the mass of total carbon emission in per unit cement production, unit: t C per t cement).

### 2. Apparent energy consumption calculation:

The activity data can be directly extracted as the final energy consumption from energy statistics, or estimated based on the mass balance of energy, the so-called apparent energy consumption estimation method:

Apparent energy consumption= domestic production + imports – exports – change in stock – non energy use of fuels (5)

Notably that the non-energy use of fossil fuels and other industrial processes such as ammonia production, lime production and steel production will also produce carbon emissions[30]. To be consistent with the scope of the international data set we are comparing, those emissions are not included in this study. Previous study suggested that such emission is equivalent to about 1.2% of China's total emissions from fossil fuel combustion and cement production process[16]

# Supplementary Materials

# New dynamics of energy use and carbon emissions in China

SI materials including 2 Figures and 4 Tables.

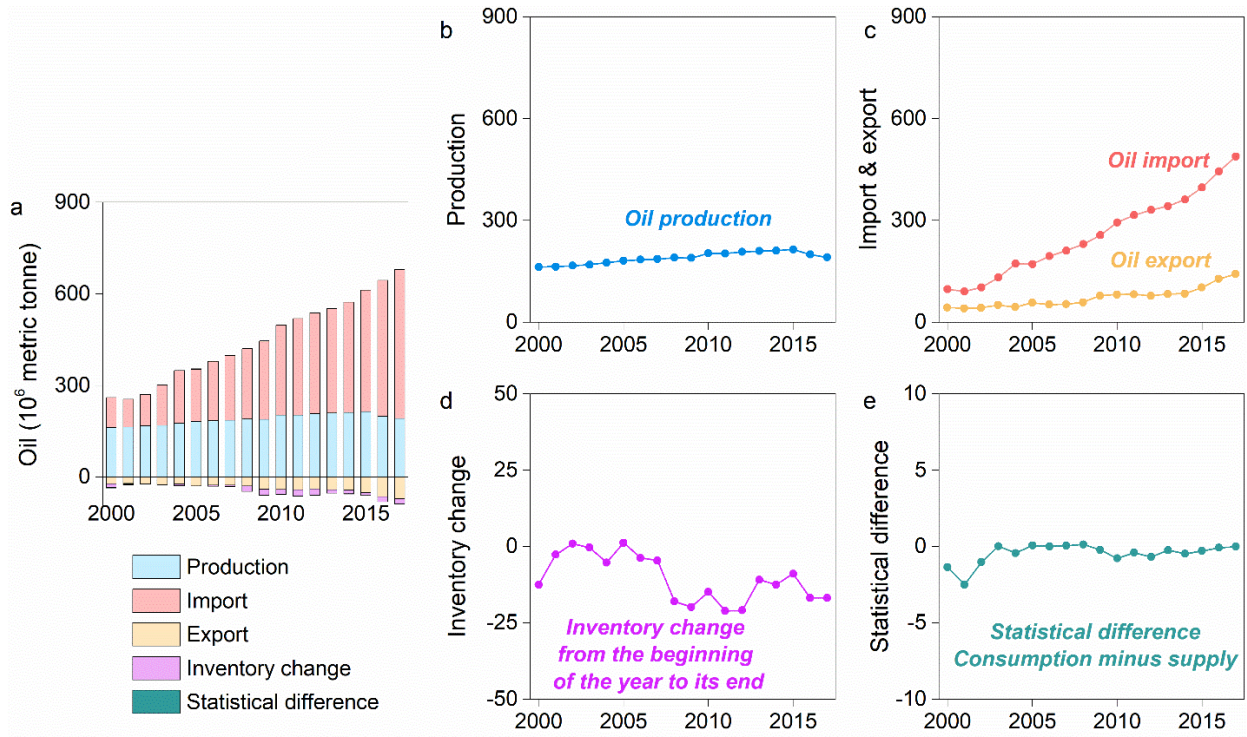

**SI Figure 1, Apparent energy consumption data for oil (2000-2018). Figure 1b, 1c, 1d and 1e share the same unit with Figure 1a: ($10^6$ metric tone)**

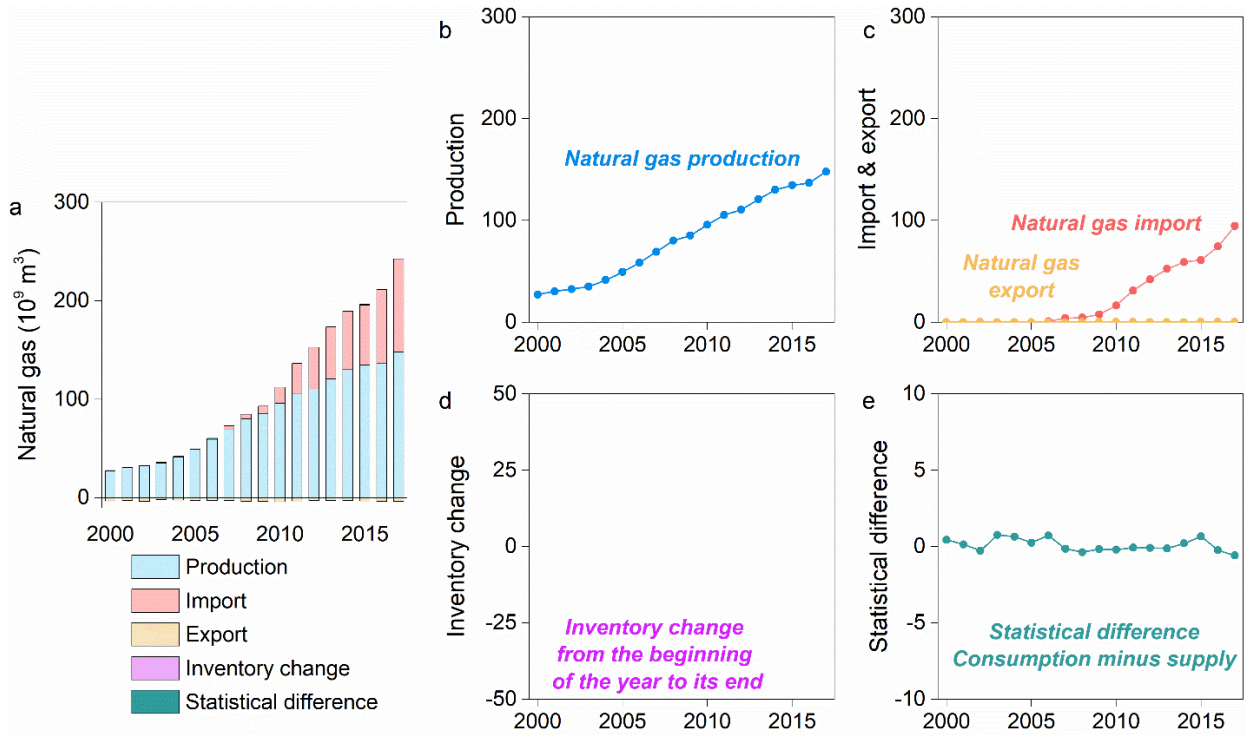

**SI Figure 2** Apparent energy consumption data for natural gas (2000-2018). SI Figure 1b, 1c, 1d and 1e share the same unit with SI Figure 1a: ($10^9$ m$^3$)

SI Table 1 Contributions of different factors to uncertainty in total $CO_2$ emissions

|  | Coal | Oil | Natural gas | Cement |
|---|---|---|---|---|
| **Production** | 18.0% | 0% | 0% | 0.1% |
| **Import** | 12.4% | 10.6% | 0.5% | / |
| **Export** | 0% | 0.5% | 0% | / |
| **Inventory change** | 26.2% | 1.8% | 0% | / |
| **Statistical errors** | 29.9% | 0% | 0% | / |
| **Total** | 86.5% | 12.9% | 0.5% | 0.1% |

SI Table 2 The Raw Coal's heating values by major coal production countries reported by UN[8]. The table suggested that the heating value of China's raw coal is significantly lower than other coal production countries.

| Country | (Unit: GJ/t) |
| --- | --- |
| China (the measured value by Liu et al., 2015[11]) | 20.95 |
| China (United Nations reported data values) | 21.4 |
| Global average | 29.3 |
| Argentina | 30.15 |
| Brazil | 30.54 |
| Canada | 28.30 |
| France | 27.05 |
| Germany | 26.65 |
| Hungary | 29.70 |
| Italy | 27.33 |
| Japan | 25.91 |
| Netherlands | 25.39 |
| New Zealand | 28.30 |
| Norway | 28.10 |
| Pakistan | 29.10 |
| Russia | 25.07 |
| Spain | 24.13 |
| Sweden | 28.86 |
| Switzerland | 28.10 |
| Turkey | 27.04 |
| United Kingdom | 25.70 |
| United States | 26.07 |
| Uruguay | 29.31 |

SI Table 3 Oxidation rate for coal by different data sources.

| Organization | Oxidation rate（100%） |
|---|---|
| This study | 92 |
| China's UNFCCC report by NDRC, 2013) | 94 |
| CDIAC (CDIAC, 2017) | 98 |
| EDGAR (European Commission, 2018)[10] | 100 |
| BP (BP, 2018)[9] | 100 |
| World Bank (World Bank, 2018)[27] | 100 |
| IEA (IEA, 2018)[28] | 98 |
| EIA (EIA, 2018)[29] | 100 |

SI Table 4 Projected growth rate of China's CO2 emissions in 2018 compared to 2017

|  | **Based on first nine months data** | **Based on first ten months data** |
| --- | --- | --- |
| **Emissions from:** |  |  |
| Coal | + 4.5% (range: 1.1% to 8.0%) | + 4.8% (range: 1.8% to 7.8%) |
| Oil | + 3.6% (range: -1.9% to 9.1%) | + 5.6 % (range: 1.3% to 9.9%) |
| Natural gas | + 17.7% (range: 14.2% to 21.2%) | + 17.4% (range: 14.2% to 20.6%) |
| Cement | + 1.0% (range: -0.6% to 2.6%) | + 2.6% (range: 1.4% to 3.8%) |
| **Total emissions** | + 4.8% (range: 1.3% to 8.3%) | + 5.5% (range: 2.5% to 8.5%) |